\documentclass[%
superscriptaddress,
reprint,
nofootinbib,
amsmath,amssymb,amsbsy,aps,
pra,
longbibliography]{revtex4-1}

\usepackage[normalem]{ulem}
\usepackage{bbm}
\usepackage{verbatim}

\usepackage{cleveref}
\usepackage{braket}
\usepackage[pdftex]{graphicx}
\usepackage{epstopdf}
\usepackage{rotating}
\usepackage{dcolumn}
\usepackage{bm}
\usepackage{mathrsfs}
\usepackage{color}
\usepackage{leftidx}
\usepackage{bbm}
\usepackage{verbatim}
\usepackage{soul,xcolor}
\usepackage[toc,page]{appendix}
\usepackage{comment}
\usepackage{multirow} 
\usepackage[caption=false]{subfig}
\usepackage{extarrows}
\definecolor{greens}{rgb}{0,0.7,0}
\usepackage{amsmath}

\usepackage{lipsum}

\newcommand{\beginsupplement}{%
        \setcounter{table}{0}
        \setcounter{section}{0}
        \renewcommand{\thetable}{S\arabic{table}}%
        \setcounter{figure}{0}
        \renewcommand{\thefigure}{S\arabic{figure}}%
     }

\newcommand{\tz}[1]{{\tau}^{z}_{(#1)}}
\newcommand{\tx}[1]{{\tau}^{x}_{(#1)}}
\newcommand{\ty}[1]{{\tau}^{y}_{(#1)}}

\begin{document}

\title{\textcolor{black}{Distinguishing dipolar and octupolar quantum spin ices using contrasting magnetostriction signatures}}
\author{Adarsh S. Patri}
\affiliation{Department of Physics and Centre for Quantum Materials, University of Toronto, Toronto, Ontario M5S 1A7, Canada}

\author{Masashi Hosoi}
\affiliation{Department of Physics, University of Tokyo, 7-3-1 Hongo, Bunkyo, Tokyo 113-0033, Japan}

\author{Yong Baek Kim}
\affiliation{Department of Physics and Centre for Quantum Materials, University of Toronto, Toronto, Ontario M5S 1A7, Canada}

\date{\today}

\begin{abstract}

Recently there have been a number of experiments on Ce$_2$Zr$_2$O$_7$ and Ce$_2$Sn$_2$O$_7$,
suggesting
that these materials host a three-dimensional quantum spin liquid with
emergent photons and fractionalized spinon excitations. 
However, the interpretation of the data to determine the 
precise nature of the quantum spin
liquids is still under debate. 
The Kramers doublet in Ce$^{3+}$ local moment
offers unusual pseudo-spin degrees of freedom as the $x$ and $z$ components
transform as a dipole and $y$ component as an octupole. 
This leads to a variety of possible quantum spin liquid (or quantum spin ice) phases on the
pyrochlore lattice of Ce$^{3+}$ moments. 
In this work, we theoretically
propose that magnetostriction would be able to distinguish
the dipolar (D-QSI) and octupolar (O-QSI) quantum spin ice, where 
the dipolar or octupolar components possess the respective spin ice correlations. 
We show that the magnetostriction in various configurations can be used as a
selection rule to differentiate not only D-QSI and O-QSI, but also a number
of competing broken symmetry states.

\end{abstract}

\maketitle


The inclusion of multipolar moments in geometrically frustrated lattices yields a diverse range of emergent phases of matter.
Multipolar moments, which characterize asymmetric charge and magnetization densities \cite{multupole_rev_1,multupole_rev_3}, typically arise from spin-orbit coupled systems subject to strong crystalline electric fields, and as such they transform non-trivially under spatial symmetries.
In the archetypal three-dimensional frustrated pyrochlore lattice, the interactions between these moments can give rise to  unusual broken-symmetry `hidden' phases (aptly named due to their shyness to common experimental probes), or a long-range entangled U(1) quantum spin liquid known as quantum spin ice \cite{hermele_pyrochlore_photons, rau_gingras_frustration,balents_sl}.
Quantum spin ice may also be understood in a relatively simpler fashion, namely the manifestation of the coherent superposition of the classical two-in, two-out degenerate manifold found in classical spin ice \cite{Gingras_2014_qsi_review,gingras_bramwell_science_review}.
Although multipolar moments are not \textit{a priori} required for the emergence of quantum spin ice, as pure dipolar systems have also been examined \cite{isakov_ybk_hcb_2008, savary_gmft,shannon_seeing_the_light, onoda_1, onada_qmc}, the higher-rank multipolar based systems offer an arguably richer diversity of possible QSLs that inherit the non-trivial symmetry properties of the underlying moments.

The dipolar-octupolar (DO) Kramers compounds, Ce$_2$(Sn,Zr)$_2$O$_7$ and Nd$_2$Zr$_2$O$_7$\cite{gang_chen_prl, ce_sn_2015, gaulin_do, nat_phys_ce, gang_chen_prr}, are such examples, where interacting dipoles and octupoles permit the existence of so-called dipolar-quantum spin ice (D-QSI) or octupolar-quantum spin ice (O-QSI), which are coherent superpositions of the `two-in, two-out' configurations of dipolar or octupolar moments, respectively.
The difference in their microscopic origin leads to their emergent gauge fields inheriting the symmetry properties of their respective multipolar moment, namely the O-QSI (D-QSI) emergent electric field transforms as an octupolar (dipolar) moment.
This leads to striking inelastic neutron scattering (INS) predictions \cite{gang_chen_prb}, where O-QSI is expected to have intensity contributions solely from the spinons, as the emergent photon cannot symmetry-permitting couple to the dipole moment of the incident neutrons.
A further diversity is in the flux configuration of the emergent vector potential, namely 0-flux or $\pi$-flux through each pyrochlore hexagon, which leads to so-called un-frustrated or frustrated quantum spin ices, which maintain or enlarge the unit cell, respectively. 

Experimental investigations into the Ce-based candidate materials rule-out pure classical-spin ice behaviours and hint at the importance of quantum fluctuations.
Indeed, recent INS measurements \cite{gaulin_do} in Ce$_2$Zr$_2$O$_7$ suggest the existence of D-QSI, as the measured intensity qualitatively agrees with predictions of contributions arising solely from low-energy photon excitations.
A concurrent experimental study \cite{nat_phys_ce} suggests Ce$_2$Zr$_2$O$_7$ does not belong to the spin ice regime, and hints at the possibility of a frustrated ($\pi$-flux) QSI, but stops short of being able to identify it as dipolar or octupolar.
What is lacking is a clear smoking-gun signature that allows the differentiation of the two types of quantum spin ice (even within the un-frustrated sector) of DO systems.

Motivated by recent lattice-based studies of QSI in non-Kramers pyrochlore materials \cite{nan_tang_aps_2019, patri2019probing}, \textcolor{black}{as well as in heavy fermion compounds \cite{ce_hfm, ce_hfm_2} and pressurized Kitaev materials \cite{magneto_kitaev}}, we theoretically propose that magnetostriction is an ideal probe to identify and isolate the two different \textcolor{black}{zero-flux} QSIs in DO systems. 
Our findings of the distinguishing signatures are based on classical analyses and exact diagonalization (ED) of quantum models on the pyrochlore lattice.
The ED studies indicate an enhancement of the quantum fluctuations about the classical solutions for both proposed QSIs.
Our predictions form the basis to clearly differentiate the two QSIs and help to advance the study of multipolar based quantum spin liquids.

\begin{figure}[t]
  \centering
  \includegraphics[width=0.8\linewidth]{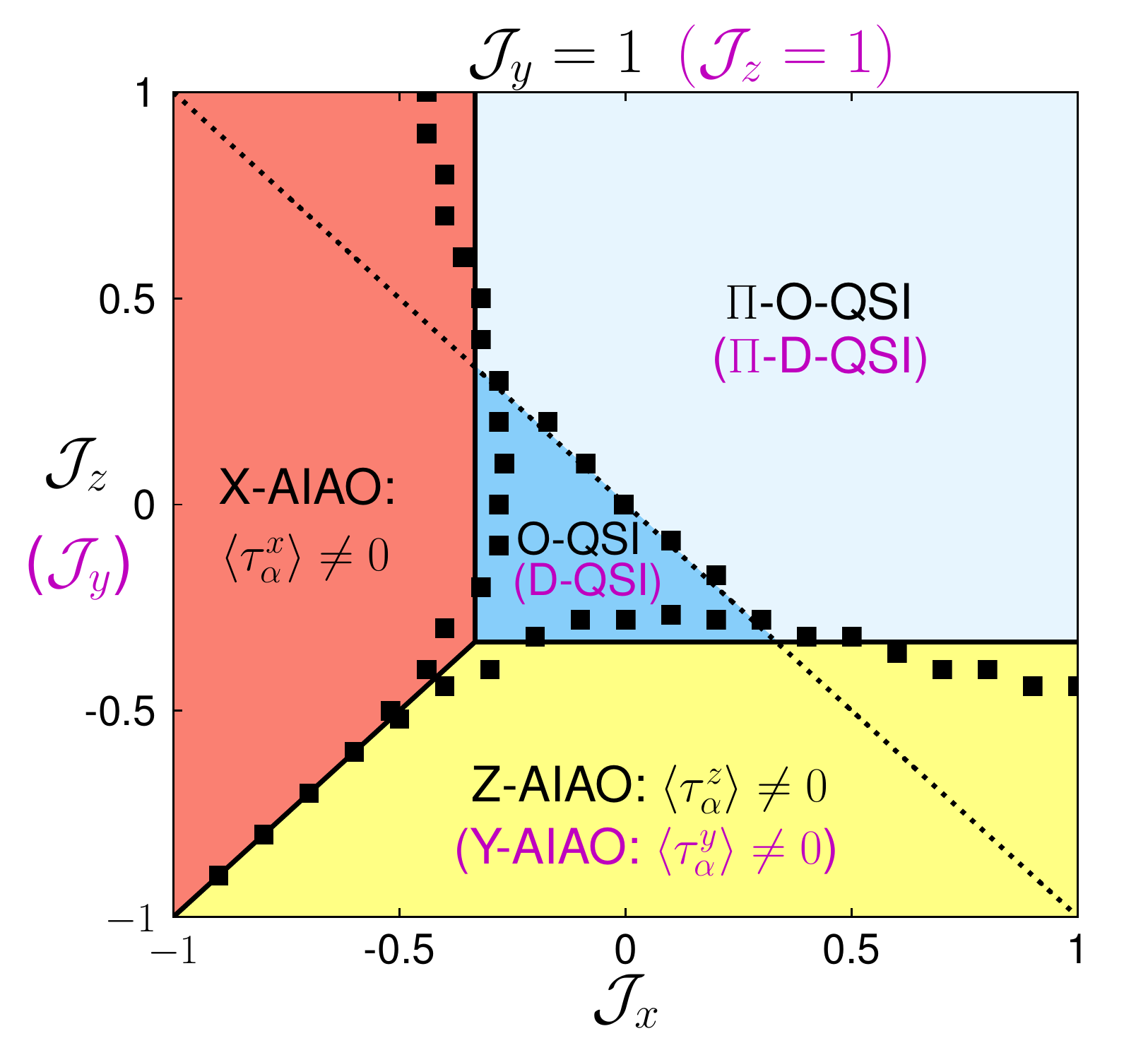}
  \caption{Phase diagram of DO model Eq. \ref{eq_XYZh} in zero magnetic field, $\bm{h} = \bm{0}$ in the $\mathcal{J}_y=1$ ($\mathcal{J}_z=1$) plane.
  The depicted phases are the 0-flux octupolar (dipolar) quantum spin ice O-QSI (D-QSI), frustrated $\pi$-flux octupolar (dipolar) quantum spin ice $\Pi$-O-QSI ($\Pi$-D-QSI), pseudospin-$\mu$ all-in, all-out ordered phase ($\mu=x$, X-AIAO; $\mu=y$, Y-AIAO; $\mu=z$, Z-AIAO).
The solid lines denote classical phase boundaries, while the black squares denote phase boundaries obtained from 16-site ED studies.
The dashed $\mathcal{J}_x+\mathcal{J}_{y,z}=0$ line denotes the crossover from un-frustrated to frustrated exchange couplings. 
We emphasize that the terminology of `dipolar' is used for due to the $\Gamma_2$ octupole transforming identically to the $J_z$ dipole ($\Gamma_2$), despite the `dipole' phases being composed of identically transforming $J_z$ dipole and $(J_x^3-\overline{J_xJ_yJ_y})$ octupole.}
  \label{fig_phasediagram}
\end{figure} 

\textit{Microscopic picture. ---}
In the pyrochlore materials, A$_2$B$_2$O$_7$, the rare-earth ion (A) is subject to a local $D_{3d}$
point group symmetry instilled by the crystalline electric field (CEF) of the surrounding oxygen cage.
Focussing on A ions carrying an odd number of electrons, this CEF splits the spin-orbit coupled degenerate $J$ manifold to yield low-lying Kramers doublet ground states, whose degeneracy is protected by time-reversal symmetry \cite{ross_savary_prx}. 
The Kramers ground states can be divided into doublets formed from a two-dimensional irreducible representation (irrep), or formed from two one-dimensional irreps of $D_{3d}$.
The first type, the more familiar Kramers ions, is found in (Yb,Er)$_2$Ti$_2$O$_7$ \cite{savary_order_by_disorder, Yasui_ybto_2003, Chang_ybto_2012, Armitage_ybto_2015, gaudet_ybto_2016, broholm_ybto_111_2017, Broholm_ybto_mm_2019, Hodges_2001, Yaouanc_ybto_2016, Armitage_ybto_2014, thompson_ybto_2017}, where the ground states transform as the two-dimensional $\Gamma_4$ irrep, and as such host conventional magnetic dipole moments, $\bm J$.
The second non-trivial type is known as a dipolar-octupolar (DO) doublet, arising in Ce$_2$(Sn,Zr)$_2$O$_7$ \cite{nat_phys_ce, gang_chen_prl,gaulin_do,ce_sn_2015}and Nd$_2$(Ir,Zr)$_2$O$_7$ \cite{nd_zr_nakatsuji,Watahiki_2011_nd_it,Balakrishnan_nd_zr,benton_nd, petit_nd, nd_zr_cef_states}, which transform as the one-dimensional irreps $\Gamma_5^+$ and $\Gamma_6^+$, respectively. 
By considering the irrep product formed by this doublet subspace, $ (\Gamma_5^+\oplus\Gamma_6^+)\otimes(\Gamma_5^+\oplus\Gamma_6^+)= 2\Gamma_1^+\oplus2\Gamma_2^+ $, the active multipoles supported by the DO ground state are found to be the $J_z$ ($\Gamma_2$) magnetic dipole, $J_x^3-\overline{J_xJ_yJ_y}$ ($\Gamma_2$) magnetic octupole, and $J_y^3-\overline{J_yJ_xJ_x}$
($\Gamma_1$) magnetic octupole, where the overline indicates a symmetrized product.
Based on the nature of the ground state wave functions, these moments can be efficiently represented in terms of a pseudospin-1/2 operator, $\bm{ \mathcal{S} }$: 
$\mathcal{S}^x = {\mathcal{C}}_0(J_x^3-\overline{J_xJ_yJ_y})+{\mathcal{C}}_1J_z$, $ \mathcal{S}^y = {\mathcal{C}}_2(J_y^3-\overline{J_yJ_xJ_x})$, and $ \mathcal{S}^z = {\mathcal{C}}_3J_z$.
Here, each $\mathcal{C}_{0,1,2,3}$ coefficient can be determined by the experimentally found CEF parameters; for Ce$_2$(Sn,Zr)$_2$O$_7$, $\mathcal{C}_1 = 0$.

\textit{DO models. ---}
From the symmetry requirements dictated by the pyrochlore lattice, the nearest-neighbour pseudospin Hamiltonian for the DO system under the influence of an applied magnetic field, $\bm{h}$, can be written in the form,
\begin{align}
\label{eq_XYZh}
  H_{\mathrm{XYZh}} &=\sum_{\langle i,j\rangle} \mathcal{J}_\mu {\tau}_i^\mu {\tau}_j^\mu -\sum_i \Bigg[({\bm h}\cdot\hat{z}_i)  \left({g}_x {\tau}_i^x+ {g}_z {\tau}_i^z \right) \Bigg. \\
  \Bigg. & ~~~~~~~~~ ~~~~~ ~~~~~ ~~~~~~~~ + g_y  \Big( (h_i^{y})^3  - 3 (h_i ^{x})^2 h_i^{y}  \Big) \tau^y_i \Bigg] \nonumber ,
\end{align}
where we employ Einstein summation notation for $\mu = \{x,y,z\}$, define a new pseudospin-1/2 operator $\bm{\tau}$, and $\hat{z}_i$ is the local-$z$ direction of the sublattice at site $i$.
It suffices to state here that $\tau^y = \mathcal{S}^y$, while ${\tau}^x = \cos(\theta) {\mathcal{S}}^x - \sin(\theta) {\mathcal{S}}^z$ and ${\tau}^z = \sin(\theta) {\mathcal{S}}^x + \cos(\theta) {\mathcal{S}}^z$, where $\theta$ is defined in terms of the exchange coupling constants of the `standard' DO model (as described in Supplementary Materials (SM) \cite{supp_mat}).
Importantly, both $\tau^{x,z}$ couple to the local-$h^z_{i}$, as both pseudospins transform identically.

We present in Fig. \ref{fig_phasediagram} the phase diagram corresponding to Eq. \ref{eq_XYZh} with $\bm{h}=\bm{0}$ for the dipolar-dominant ($\mathcal{J}_z=1$) and octupolar-dominant ($\mathcal{J}_{y}=1$) regimes.
16-site ED phase boundaries are overlapped on top of the classical phase diagrams.
As seen, the ED phase boundaries agree very well with the classical transition lines in both dipole and octupole dominant regimes.
Moreover, the location of the obtained ED phase boundaries are faithfully comparable to those obtained from parton (gauge) mean-field theory studies \cite{gang_chen_prl}.
We employ the terminology of `dipole' dominant phases, despite being formed from $J_z$ dipole and $(J_x^3-\overline{J_xJ_yJ_y})$ octupole, as both microscopic moments transform as the $J_z$ dipole.
For each subfigure, there exists an unfrustrated 0-flux quantum spin ice such as D-QSI (O-QSI), and a frustrated $\pi-$flux quantum spin ice $\Pi$-D-QSI ($\Pi$-O-QSI); we describe the unfrustrated phases shortly.
These different-flux phases are classically indistinguishable; however in the 16-site ED, non-analytic signatures in the ground-state energy indicate a phase boundary separating them, as seen in Fig. \ref{fig_phasediagram}(a),(b).
This further highlights the importance of the 16-site ED results.
We henceforth focus on the unfrustrated regimes of the model i.e. where $\mathcal{J}_{x} + \mathcal{J}_{y} <0$ and $\mathcal{J}_{x} + \mathcal{J}_{z} <0$, as we are interested in examining the $\bm{q}=\bm{0}$ pseudospin orderings' magnetostriction behaviours.
The two un-frustrated varieties of spin ices for the DO system are the aforementioned D-QSI (O-QSI), which in the classical limit corresponds to a degenerate manifold of two-in, two-out $\tau^z$ ($\tau^y$) moments. 
There also exist a variety of broken-symmetry all-in, all-out (AIAO) dipolar and octupolar phases where we have uniform pseudospin ordering on each sublattice: X-AIAO, Y-AIAO, and Z-AIAO which corresponds to AIAO ordering of $\tau^x$, $\tau^y$ and $\tau^z$ moments, respectively.

\textit{Lattice-pseudospin couplings. ---}
Due to the time-reversal odd nature of dipoles and octupoles, they can only couple to the lattice degrees of freedom with the assistance of an external magnetic field, $\bm{h}$.
Imposing the underlying local $D_{3d}$ constraint of the surrounding cage (as described in SM \cite{supp_mat}), the dipolar and octupolar moments couple to the elastic strain as,
\begin{widetext}
\begin{align}
\mathcal{F}_{DO} &= \tau^x \Bigg[ g_1 \left( h^x _{\alpha} \Big(\epsilon_{yy }^{\alpha} - \epsilon_{xx} ^{\alpha} \right)  + 2 h^y_{\alpha} \epsilon_{xy}^{\alpha}  \Big)  + g_2 \left( h^x _{\alpha} \epsilon_{xz}^{\alpha} + h^y _{\alpha} \epsilon_{yz}^{\alpha} \right) + g_3 h^z _{\alpha} \left(\epsilon_{xx}^{\alpha} + \epsilon_{yy}^{\alpha} \right) + g_4 h^z _{\alpha} \epsilon_{zz} ^{\alpha}  \Bigg] \nonumber \\
& + \tau^y \Bigg[  	g_5 \Big(2 h^x _{\alpha} \epsilon_{xy}^{\alpha} + h^y _{\alpha} \left(\epsilon_{xx} ^{\alpha} - \epsilon_{yy} ^{\alpha} \right)	\Big) + g_6 \Big(  h^y _{\alpha} \epsilon_{xz}^{\alpha} - h^x _{\alpha} \epsilon_{yz}^{\alpha}	\Big)	\Bigg]  \label{eq_elastic_spin_coupling} \\
& + \tau^z \Bigg[ g_7 \left( h^x _{\alpha} \Big(\epsilon_{yy }^{\alpha} - \epsilon_{xx} ^{\alpha} \right)  + 2 h^y_{\alpha} \epsilon_{xy}^{\alpha}  \Big)  + g_8 \left( h^x _{\alpha} \epsilon_{xz}^{\alpha} + h^y _{\alpha} \epsilon_{yz}^{\alpha} \right)  + g_9 h^z _{\alpha} \left(\epsilon_{xx}^{\alpha} + \epsilon_{yy}^{\alpha} \right) + g_{10} h^z _{\alpha} \epsilon_{zz} ^{\alpha}  \Bigg] \nonumber
\end{align}
\end{widetext}
where we have introduced Einstein summation notation for the sublattice index $\alpha = {0,1,2,3}$, and $g_{1, ... 10}$ are phenomenological coupling constants. 
The $\alpha$-superscript (subscript) on the magnetic field (elastic strain) denotes the respective quantities in the local coordinate system of sublattice $\alpha$. 
We highlight that $\tau^x$ and $\tau^z$ couple identically to the elastic strain, as they both transform as the basis functions of $\Gamma_2^+$ irrep.
We discuss in SM \cite{supp_mat} how the the lattice-pseudospin coupling in Eq. \ref{eq_elastic_spin_coupling} results in distorting the elastic normal modes of the pyrochlore lattice to yield the length change, $(\frac{\Delta L}{L})^{\hat{n}}_{\bm{\ell}}$, where the subscript and superscript $\bm{\ell}$ and $\bm{\hat{n}}$  label the direction of length change and applied magnetic field direction, respectively.

\begin{figure*}[t]
  \centering
  \includegraphics[width=0.9\linewidth]{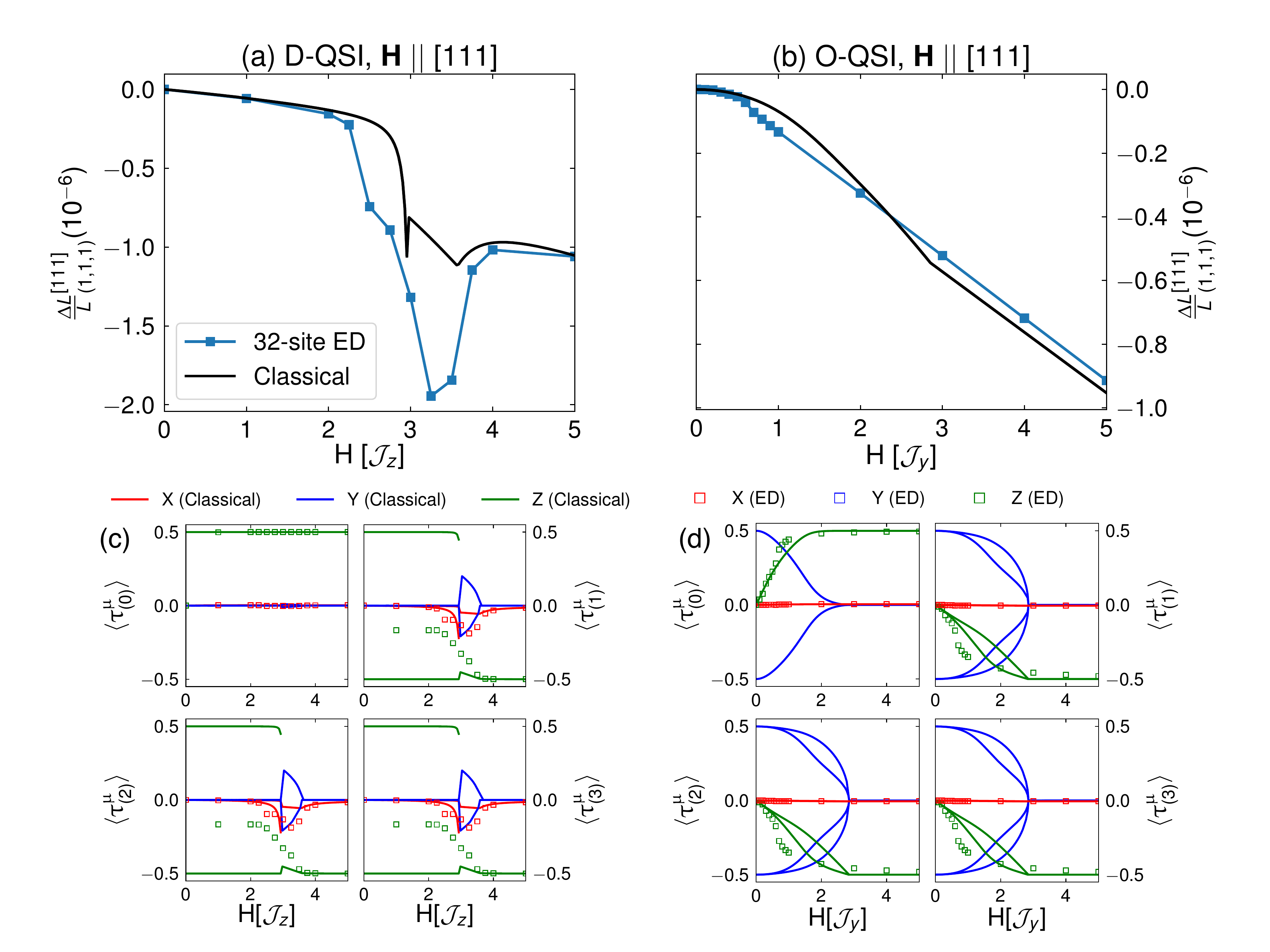}
  \caption{\textcolor{black}{Magnetostriction along $\bm{\ell} = (1,1,1)$ direction and order parameter evolution under $\bm{\hat{n}} = [111]$ magnetic field for} dipolar quantum spin ice (D-QSI) where the spin ice correlations are between pseudospin-Z moments, and octupolar quantum spin ice (O-QSI) where the spin ice correlations are between pseudospin-Y moments.
\textcolor{black}{(a), (b): Relative length change, $\frac{\Delta L}{L}$, for D-QSI and O-QSI, respectively, along $\bm{\ell} = (1,1,1)$ direction and $\bm{\hat{n}} = [111]$ magnetic field.}
  Solid lines (squares) indicate classical (32-site exact diagonalization) magnetostrictions and order parameters.
  The superimposed \textcolor{black}{magnetostriction} ED result in \textcolor{black}{(a)} indicates an enhancement by quantum fluctuations.
\textcolor{black}{(c): D-QSI order parameter evolution.} D-QSI develops (both classically and quantum mechanically) into the Kagome ice (KI) phase in the low field limit.
 Upon increasing the field, the KI undergoes a meta-magnetic transition in $\langle \tau^z \rangle $ and is accompanied by an `island' of finite $\langle \tau^{x,y}_{(1,2,3)} \rangle$ that survives for a small window of magnetic field strengths. The first (second) discontinuity \textcolor{black}{in (a)} reflects the appearance (disappearance) of this `island'.  
\textcolor{black}{(d): O-QSI order parameter evolution.} O-QSI steadily collapses with increasing field strength, and is accompanied by the gradual increase of $\langle \tau^z \rangle $ into the fully polarized phase. The single classical `kink' at $h \approx 3 \mathcal{J}_{y}$ \textcolor{black}{in (b)} is the critical field value where $\langle \tau^y \rangle$ on all sublattices has collapsed to zero \textcolor{black}{(see SM for expanded discussion on order parameters \cite{supp_mat})}.}
  \label{fig_qsi_111}
\end{figure*} 

\textit{Magnetostriction of O-QSI and D-QSI under [111] magnetic field. ---}
By considering the magnetic field applied along the $\bm{\hat{n}} = [111]$ direction, the parallel direction magnetostriction behaviour is,
\begin{widetext}
\begin{align}
\left( \frac{\Delta L}{L} \right)^{[111]}_{(1,1,1)} &= {h} \Bigg[  \mathcal{G}_z \left( 3 \tz0 - \tz1 - \tz2 - \tz3 \right) +\mathcal{G}_x \left( 3 \tx0 - \tx1 - \tx2 - \tx3 \right)  +   \mathcal{W}_x  \left(\tx1 + \tx2 + \tx3 \right) \Bigg. \nonumber \\
\Bigg. & ~~~~ ~~~  +  \mathcal{W}_z \left(\tz1 + \tz2 + \tz3 \right) + \mathcal{V}_x \left(9 \tx0 + \tx1 + \tx2 + \tx3 \right) + \mathcal{V}_z \left(9 \tz0 + \tz1 + \tz2 + \tz3 \right) \Bigg],
\end{align}
\end{widetext}
where the re-defined the pseudospin-elastic coupling constants ($\mathcal{G}_{x,z},\mathcal{V}_{x,z},\mathcal{W}_{x,z}$) are given in SM \cite{supp_mat}.
Figure \ref{fig_qsi_111} depicts the classical and quantum magnetostriction behaviours of the two spin ices parallel to a [111] magnetic field, \textcolor{black}{along with their respective order parameter evolutions.} 
\textcolor{black}{
The 32-site ED calculation requires 640 cores ($\sim 17$ hours) for magnetic field sweeps over a single parameter set; details of the ED method are given in SM \cite{supp_mat}.
We estimate the magnitude of the coupling constants $g_{1,...10}$ from comparison to magnetostriction experiments on familial rare-earth Pr- and Ce- based heavy fermion compounds \cite{gegenwert_pr,ce_hfm, ce_hfm_2}.
In these materials, the measured relative length changes are on the order of $10^{-6}$, which can be achieved here by the numerical values presented in SM \cite{supp_mat}.
The precise values of the coupling constants can be determined in employing our theoretical predictions in conjunction with (future) experimental measurements on Ce$_2$(Sn,Zr)$_2$O$_7$; for example, by fitting the experimentally measured length changes along the various directions (and field orientations) proposed in SM \cite{supp_mat}.}
As seen in Fig. \ref{fig_qsi_111}(a), (b), there is a clear discrepancy between the two spin ices; we present the unique evolution of the respective order parameters in SM \cite{supp_mat}.

The classical D-QSI experiences a sharply decreasing jump-discontinuity at $h_a \approx 3 \mathcal{J}_{z}$, followed by a `drop', and then a sharp kink in the length change at $h_b \approx 3.5 \mathcal{J}_{z}$, \textcolor{black}{as seen in Fig. \ref{fig_qsi_111}(a)}.
From the quantum mechanical ED results, the D-QSI is in agreement with its classical result in that it displays a similar `drop' around the same magnetic field strength.
The magnitude of the `drop', however, is enhanced in the ED as compared to its classical counterpart, which indicates the importance of quantum fluctuations in enhancing D-QSI's magnetostriction.
The underlying physics of the D-QSI magnetostriction behaviour can be understood in terms of a meta-magnetic transition from Kagome ice (classically two-in, two-out, with sublattice-0 fixed to $+1/2$) to the fully polarized (three-in, one-out) phase, \textcolor{black}{as seen in the corresponding order parameter evolution in Fig. 2(c)}.
This transition is accompanied by the brief appearance and disappearance of an `island' of $\langle \tau^{x,y} \rangle \neq 0$, which accounts for the aforementioned sharp non-analytic `kinks' in Fig. \ref{fig_qsi_111}(a) at $h_a$, $h_b$. In the quantum model, the meta-magnetic transition is more gradual, and lacks the sharp discontinuous features of the classical island, which is expected from a finite-sized cluster, but nevertheless creates the (enhanced by quantum-fluctuations) `dip' feature in the magnetostriction.
Indeed, the analogous physics is responsible for the similar magnetostriction results proposed for non-Kramers QSI in Pr$_2$Zr$_2$O$_7$\cite{patri2019probing}.

On the other hand, the classical O-QSI, undergoes a monotonic (negative) increase in the length change with a single continuous `kink' at $h_o \approx 3 \mathcal{J}_{y}$, \textcolor{black}{as seen in Fig. \ref{fig_qsi_111}(b).}
The origin of the `kink' is the ultimate demise of pseudospin-Y, and the completed polarization of pseudospin-Z which is encouraged by the [111] magnetic field, \textcolor{black}{as seen in the order parameter evolution in Fig. \ref{fig_qsi_111}(d)}. 
Indeed as the magnitude of pseudospin-Y is diminishing with increasing field, two sublattices' ($\alpha_1$ and $\alpha_2$) pseudospin-Y expectation values are positive while the remaining two sublattices' ($\alpha_3$ and $\alpha_4$) pseudospin-Y expectation values are negative i.e. $\langle \tau^y _{(\alpha_1, \alpha_2)} \rangle > 0$, while on the other two sublattices $\langle \tau^y _{(\alpha_3, \alpha_4)} \rangle < 0$.
This sign-structure is reminiscent of the octupolar spin ice two-in, two-out degeneracy. 
{The increasing pseudospin-Z in conjunction with the disappearing pseudospin-Y thus indicates 
a polarized dipole (pseudospin-Z) coexisting with octupole (pseudospin-Y) spin-ice correlations
for $h \leq h_o $.
For the quantum model's magnetostriction, there is an overall monotonic (negative) increase in the length change that is analogous to the classical behaviour.
Interestingly, \textcolor{black}{as seen in Fig. \ref{fig_qsi_111}(b)}, the classical `kink' at $h_o$ is smoothened out in the quantum model, which can be understood from the ED order parameter evolution under the magnetic field: the transition into the fully-polarized state is smoother/broadened out in the ED order parameters \textcolor{black}{in Fig. 2(d)}. 
Moreover, there exist non-analytic kinks in the order parameter (and the second derivative of the ED ground state energy, SM \cite{supp_mat}) at an order of magnitude smaller than the classical $h_o$.
The early locations of the ED `kinks' suggest the fragility of the O-QSI to quantum fluctuations in the presence of the field.}

The key difference between D-QSI and O-QSI is with the ability of the pseudospin degree of freedom responsible for forming the classical ice manifold to couple directly to the magnetic field.
(SM \cite{supp_mat} contains the magnetic field couplings in a variety of magnetic field directions for the two spin ices.)
This leads to different phases appearing in the low field window, and subsequently distinct magnetostriction signatures of the parent spin ice phase.
Despite this difference, both spin ices retain remnants of their parent classical SI degeneracy in the low field window, which is reflected in particular length change directions.
We present in SM \cite{supp_mat} the classical magnetostriction behaviours under the [111] field along the (1,1,0) and (0,0,1) directions, where the retained classical degeneracy is reflected. 
Once again, the D-QSI classically demonstrates a more dramatic `peak' in the magnetostriction, while the O-QSI classically has more `kink' like features.
This is the general discriminating feature between the spin ices: O-QSI possesses gentler behaviour in its magnetostriction, when compared to the sharp features of D-QSI.
In order to contrast with (and emphasize the uniqueness of) the length change behaviours of the spin ices, we have also examined (SM \cite{supp_mat}) the magnetostriction of neighbouring multipolar ordered phases.
Considering an array of commonly accessible (in cubic materials) magnetic field and length change directions, our findings highlight the anisotropic (and distinct) nature of magnetostriction for the various possible ordered phases.

\textit{Discussions. ---}
In this work, we theoretically demonstrated that magnetostriction is a keen probe to provide distinct signatures of the two types of quantum spin ice proposed in DO pyrochlore materials.
Employing a symmetry constrained lattice-pseudospin coupling, we find that D-QSI exhibits sharper non-analytic features in the magnetostriction, than O-QSI.
In terms of future work,
it would also be interesting to examine related finite temperature behaviours of length change of the spin ices, namely thermal expansion, and elastic constant softening, which are relevant experimental tools in studies of multipolar phases.
Moreover, a recent experimental report \cite{sibille2019quantum} on Ce$_2$Sn$_2$O$_7$ suggests the possible existence of $\pi$-flux (frustrated) O-QSI, which lies very close to the 0-flux (un-frustrated) QSI.
It would be intriguing if our proposed magnetostriction behaviour can provide an insight into the nature of the phase.

\vspace{1mm}

\begin{acknowledgments}

\textit{Acknowledgements. ---}
We are grateful to Bruce Gaulin for helpful discussions about experiments on Ce$_2$Zr$_2$O$_7$.
This work was supported by NSERC of Canada, and the Center for Quantum Materials at the University of Toronto. Y.B.K. is supported by the Killam Research Fellowship of the Canada Council for the Arts.
Computations were performed on the Niagara supercomputer at the SciNet HPC Consortium, and on the Graham supercomputer of Compute Canada. SciNet is funded by: the Canada Foundation for Innovation; the Government of Ontario; Ontario Research Fund - Research Excellence; and the University of Toronto.
This work was partly performed at the Aspen Center for Physics, which is supported by National Science Foundation grant PHY-1607611.
M.H. is supported by the Japan Society for the Promotion of Science through Program for Leading Graduate Schools (MERIT) and Overseas Challenge Program for Young Researchers.

\end{acknowledgments}

\clearpage
\beginsupplement

\begin{widetext}
\section*{Supplementary Materials (SM)}
\end{widetext}

\section{XYZ model couplings relation to original DO model}

The DO model employed in the main text can be obtained from the more familiar form of pseudospin-1/2 pyrochlore models \cite{rau_gingras_frustration, gang_chen_prl, gang_chen_prb}, 
\begin{widetext}
\begin{align}
  H= & \sum_{\langle i,j\rangle} \Big[J_{xx} \mathcal{S}_i^x \mathcal{S}_j^x+J_{yy} \mathcal{S}^y \mathcal{S}_j^y+J_{zz} \mathcal{S}_i^z \mathcal{S}_j  ^z +J_{xz}( \mathcal{S}_i ^x  \mathcal{S}_j ^z+  \mathcal{S}_i^z  \mathcal{S}_j^x) \Big]  - \sum_i \Bigg[ ({\bm h}\cdot\hat{z}_i) \left( \tilde{g}_z \mathcal{S}^z_i +   \tilde{g}_x  \mathcal{S}^x_i \right) + g_y  \Big( (h_i^{y})^3  - 3 (h_i ^{x})^2 h_i^{y}  \Big) \mathcal{S}^y_i \Bigg],
  \label{eq_do_original}
\end{align}
\end{widetext}
where $\hat{z}_i$ is the local-$z$ direction of the sublattice at site $i$.
We stress here that both $\mathcal{S}^x$ and $\mathcal{S}^z$ couple linearly to the strength of the magnetic field, as they both transform identically to the $J_z$ magnetic dipole moment under the $D_{3d}$ point group (i.e. $\Gamma_2^+$ irrep).
The $J_{xz}$ term indicates a mixing between the identically-transforming $ \mathcal{S}^{x}$ and $ \mathcal{S}^z$ moments.
We can eliminate this term by performing a pseudospin $\theta$ rotation about the local $\hat{y}_i$ axis, to transform the above Hamiltonian into Eq. 1 in the main text.
This transformation results in a re-definition of the pseudospin operators and couplings,
\begin{align}
&{\tau}^x = \cos(\theta) {\mathcal{S}}^x - \sin(\theta) {\mathcal{S}}^z \nonumber \\
&{\tau}^y  = \mathcal{S}^y  \\
& {\tau}^z = \sin(\theta) {\mathcal{S}}^x + \cos(\theta) {\mathcal{S}}^z \nonumber
\end{align}
where $\tan(2\theta) = \frac{2 J_{xz}}{J_{zz} - J_{xx}}$.
As well, the new magnetic field coupling coefficients are ${g_x} = \tilde{g}_z \cos(\theta) +  \tilde{g}_x \sin(\theta)$ and ${g_z} = -  \tilde{g}_z \sin(\theta) +  \tilde{g}_x \cos(\theta)$.
Finally, exchange couplings are also renormalized to be,
\begin{align}
&\mathcal{J}_x = \frac{J_{zz} + J_{xx}}{2} - \frac{\sqrt{(J_{zz}-J_{xx})^2 + 4 J_{xz}^2}}{2} \nonumber \\
&\mathcal{J}_y = J_{yy} \\
&\mathcal{J}_z = \frac{J_{zz} + J_{xx}}{2} + \frac{\sqrt{(J_{zz}-J_{xx})^2 + 4 J_{xz}^2}}{2}\nonumber 
\end{align}
In this work, we have focused on the ${g}_x\ll {g}_z$ limit, which physically corresponds to small mixing of the octupolar and dipolar moment.
This limit permits an isolated study of the octupolar and dipolar phases.

\section{Magnetic Field coupling to multipolar moments}
\label{app_mag_coupling}

We present in Table \ref{tab_mag_field} the direct magnetic field coupling to the multipolar moments under the three considered magnetic field directions: [111], [110], and [001].
We highlight that the pure octupolar moment, $\tau^y$, only couples to the magnetic field along $\hat{{\bm{n}}} \ ||$ [110].
\renewcommand{\arraystretch}{1.1}
\begin{table}[h]
\begin{tabular}{c c|c|c|c|c}
& & \multicolumn{4}{c}{Coupling to sublattice $\alpha$}  \\
&  & 0 & 1 & 2 & 3  \\
\hline 
\multirow{2}{*}{[111]} & $\tau^y _{\alpha}$ & 0 & 0 & 0 & 0 \\
& $\tau^{x,z} _{\alpha}$ & $h$ & $-\frac{h}{3}$ & $-\frac{h}{3}$ & $-\frac{h}{3}$ \\
\hline 
\multirow{2}{*}{[110]} & $\tau^y _{\alpha}$ & 0 & $-h^3$ & $h^3$ & 0 \\
& $\tau^{x,z} _{\alpha}$ & $ \sqrt{\frac{2}{3}} h$ & $0$ & $0$ & $-\sqrt{\frac{2}{3}} h$ \\
\hline 
\multirow{2}{*}{[001]} & $\tau^y _{\alpha}$ & 0 & $0$ & $0$ & 0 \\
& $\tau^{x,z} _{\alpha}$ & $ \frac{h}{\sqrt{3}} $ & $ - \frac{h}{\sqrt{3}}$ & $ -\frac{h}{\sqrt{3}} $ & $\frac{h}{\sqrt{3}}$ \\
\end{tabular}
\caption{Direct coupling of multipolar moments to magnetic field along $\bm{h} = \frac{h}{\sqrt{3}} (1,1,1)$, $\bm{h} = \frac{h}{\sqrt{2}} (1,1,0)$, and $\bm{h} = h (0,0,1)$ directions.}
\label{tab_mag_field}
\end{table}

\begin{figure*}[t]
\includegraphics[width= 0.45\linewidth]{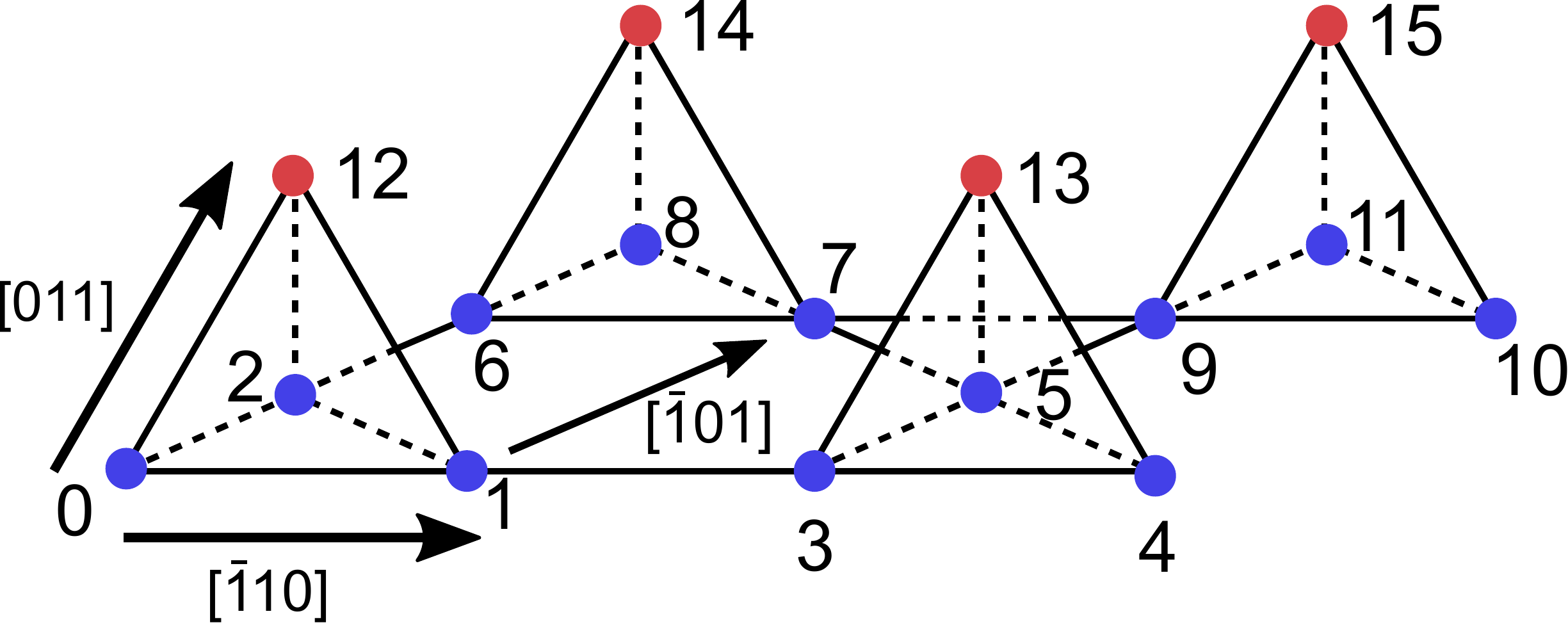} 
\includegraphics[width=0.45\linewidth]{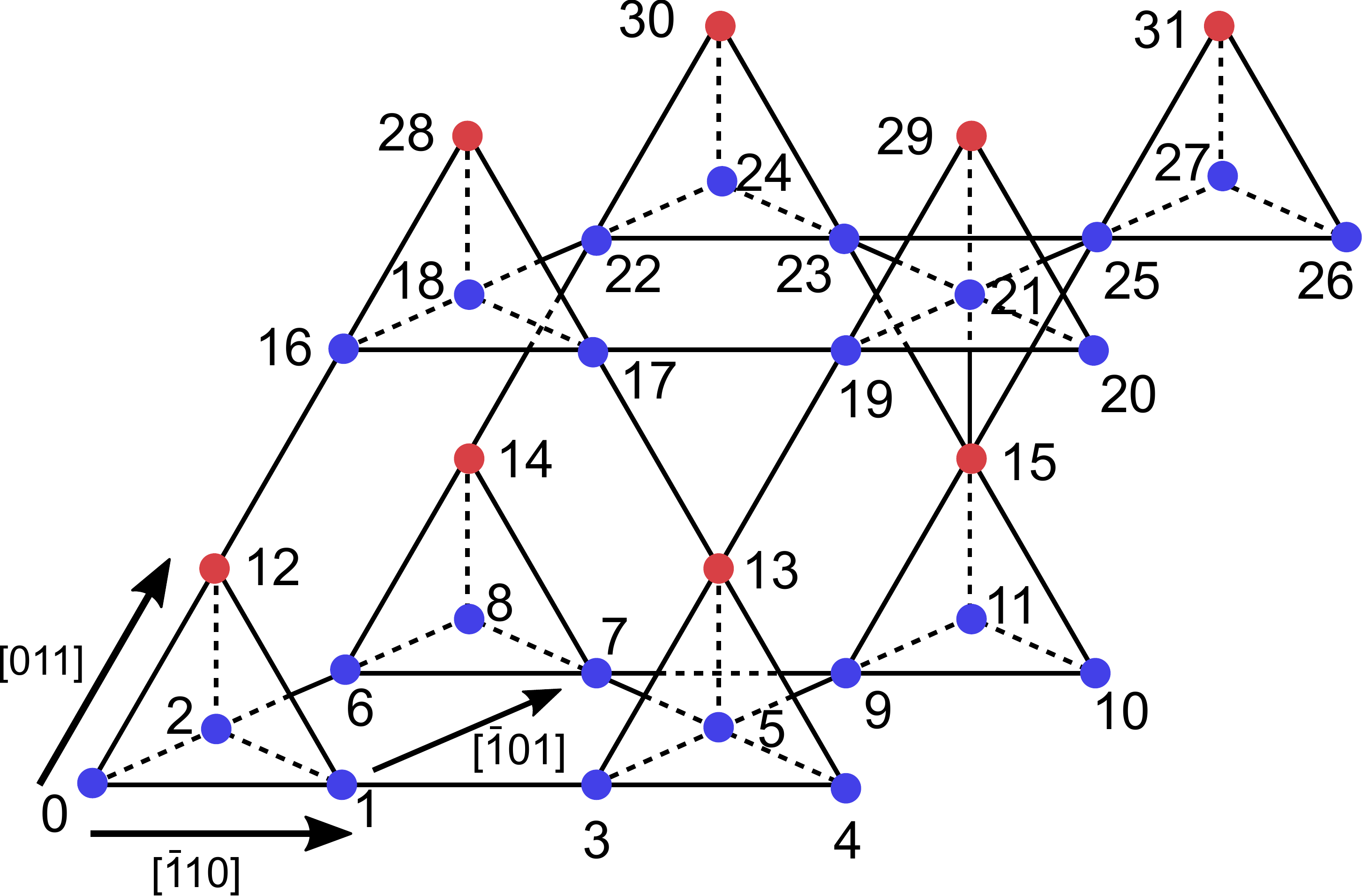} 
\caption{Finite-sized ED clusters.
Left (Right) Panel: Schematic of the 16- (32-) ED clusters, with the red and blue spheres denoting the locations of the pseudospins on the Kagome and triangular planes, respectively. }
\label{fig_ed_clusters}
\end{figure*}

\section{Symmetry Transformations of Multipolar Moments}
\label{app_symm_multipoles}

The local $D_{3d}$ point group can be generated by (i) $  \mathcal{S}_{6z} ^{-}$: $\pi/3$ improper rotation about the local $\hat{z}$-axis, and (ii)  $  \mathcal{C}_{2y}  $:  $\pi$ rotation about the local $\hat{y}$-axis. Under these generating elements, the pseudospins transform as,
\begin{align}
& \bm{\tau} \xrightarrow{\mathcal{S}_{6z} ^{-}}  \bm{\tau} \nonumber \\
& \tau^{x,z} \xrightarrow{\mathcal{C}_{2y}} - \tau^{x,z} \\
& \tau^{y} \xrightarrow{\mathcal{C}_{2y}}  \tau^{y} \nonumber
\end{align}
As seen, $\tau^x$ transforms identically to $\tau^z$ as they are both basis function of the $\Gamma_2^+$ irrep of $D_{3d}$.

\section{Length change from pseudospin-lattice coupling}
\label{app_length_change_deriv}

The underlying cubic nature of the pyrochlore lattice constrains the elastic free energy to be of the form,
\begin{equation}
\begin{aligned}
\mathcal{F}_{\text{lattice}} &= \frac{c_{B}}{2} \left(\epsilon_B^2 \right) + \frac{c_{11} - c_{12}}{2} \left( \epsilon_{\mu} ^2 + \epsilon_{\nu} ^2 \right)  \\
& + \frac{c_{44}}{2} \left( \epsilon_{xy}^2 + \epsilon_{yz} ^2 + \epsilon_{xz} ^2 \right)  ,
\label{F_lattice_normal}
\end{aligned}
\end{equation}
where the crystal's deformation is described by the strain tensor $\epsilon_{ik}$, and $c_{ij}$ is the elastic modulus tensor describing the stiffness of the crystal. Here $c_B$ is the bulk modulus,  $\epsilon_B \equiv \epsilon_{xx} + \epsilon_{yy} + \epsilon_{zz}$ is the volume expansion of the crystal, $\epsilon_{\nu} \equiv (2 \epsilon_{zz} - \epsilon_{xx} - \epsilon_{yy}) / \sqrt{3}$ and $\epsilon_{\mu} \equiv (\epsilon_{xx} - \epsilon_{yy})$ are the cubic normal mode strains.
Equipped with the couplings in Eq. 2 of the main text, we minimize $\mathcal{F}_{\text{lattice}} + \mathcal{F}_{DO}$ with respect to the cubic normal modes to obtain the extremized strain tensors, which are dependent on the pseudospin configurations and magnetic field strengths.
Inserting the extremized strain into the length change formula given by $(\frac{\Delta L}{L})^{\hat{n}}_{\bm{\ell}}=  \Sigma_{j, k= 1} ^3 \epsilon_{j k} \hat{\ell}_j \hat{\ell}_k$ yields the magnetostriction along a direction $\bm{\ell}$ in the presence of a magnetic field $\bm{h} = h \bm{\hat{n}}$.

\section{Exact Diagonalization (ED) Method}
The ED ground state is extracted by using the quantum model solver package H$\Phi$ \cite{h_phi_package}.
The eigenstates, eigenenergies, one-body Green's function, and the two-body Green's function are obtained directly from the aforementioned package.
The convergence factor of the Lanczos algorithm is set to be $10^{-9}$.
The one-body Green's function permits the extraction of the expectation value of the pseudospin operator on a given sublattice $\alpha$ i.e.  $\langle \tau _{\alpha}^\mu\rangle$.
The reason we are able to do so for particular field directions is due to the magnetic field explicitly breaking the symmetry, thus enabling finite expectation values (indicative of the symmetry broken phase) from a finite-sized cluster.
As such, for the O-QSI, for example, we cannot obtain $\tau^y$ expectation value for fields applied along the [111] and [001] directions.
Fortunately, (1,1,1) length change for the [111] magnetic field does not involve the pseudospin-Y expectation value directly, and as such the quantum magnetostriction behaviour can be extracted.

\textcolor{black}{
In the main text, we presented the order parameter evolution under an increasing [111] magnetic field.
The classically degenerate branches are apparent for both D-QSI and O-QSI in the low field regimes, which reflect the spin ice correlations in pseudospin-Z and pseudospin-Y respectively.
In particular, for the D-QSI, the two-in, two-out correlations of the parent spin ice are retained in the Kagome ice regime, which results in three classically degenerate solutions, $ \{ \langle \tau^z_{(0)} \rangle,  \langle \tau^z_{(1)} \rangle,  \langle \tau^z_{(2)} \rangle,  \langle \tau^z_{(3)} \rangle  \}= \{  \frac{1}{2}, \frac{1}{2}, -\frac{1}{2}, -\frac{1}{2} \}, \{  \frac{1}{2}, -\frac{1}{2}, \frac{1}{2}, -\frac{1}{2} \}, \{  \frac{1}{2}, -\frac{1}{2}, -\frac{1}{2}, \frac{1}{2} \}$, where the sublattice-0 is polarized with an infinitesimal field.
The corresponding ED ground state is non-degenerate, with pseudospin-Z expectation value of $-\frac{1}{6}$ on each sublattice-1,2,3.
This suggests that the quantum ground state is an equal superposition over each of the three classically degenerate manifolds.
For the O-QSI, the ED results for pseudospin-Z obey the same monotonic change as the classical results for all sublattices.
}

We present in Fig. \ref{fig_ed_clusters} the 16 and 32 site ED clusters we employ in this work.
This cluster is formed by lattice points in each of the $x$, $y$, and $z$ directions, and we impose periodic boundary conditions in the three directions.
The 16-site ED cluster has the drawback of having only one unit cell in the [011] direction.

\section{Numerical values of pseudospin-lattice and magnetic field couplings}

For the magnetostriction behaviours, we use the following strengths of the couplings,
$g_1 = 4 \times 10^{-7}$, $g_2 = -8 \times 10^{-7}$, $g_3 = 12 \times 10^{-7}$, $g_4 = -2.6 \times 10^{-7}$, $g_5 = 0.27 \times 10^{-7}$, $g_6 = -0.8 \times 10^{-7}$, $g_7 = 0.5 \times 10^{-7}$, $g_8 = -0.7 \times 10^{-7}$, $g_9 = 0.43 \times 10^{-7}$, and $g_{10} = 0.51 \times 10^{-7}$.
\textcolor{black}{As discussed in the main text, the scale of the above coupling constants results in magnetostriction behaviours on the physical scale of $(\Delta L /L) \sim 10^{-6}$.}
The order of magnitude choice of the pseudospin-X coupling terms aids in highlighting the importance of quantum fluctuations in the D-QSI and O-QSI. 
If these are chosen to be comparable, then magnetostriction is dominated by the pseudospin-Z, and gives an analogous `jump' feature to what is seen in recent magnetostriction studies of multipolar spin ice under a [111] field \cite{patri2019probing}.
We take the elastic constants as unity, $c_{B} = c_{44} = (c_{11} - c_{12}) = 1$.
For the magnetic field terms, we take $g_z = 1$, $g_x = 0.01$, and $g_y = 4 \times 10^{-4}$.
The diminutive nature of $g_y$ emphasizes the perturbatively weak nature of the cubic-in-$h$ coupling.
The minuscule nature of $g_x$ ensures very small mixing between the pure dipole ($J_z$) and octupole ($J_x^3-\overline{J_xJ_yJ_y}$), and thus allows an isolated study of the dipolar-dominant and octupolar-dominant phases.
For the D-QSI magnetostriction, {although the tiny $g_x$ coupling gives a very small  $\langle \tau^x_{(0,1,2,3)} \rangle \neq 0 $ in the fully polarized limit, we can ignore them in this discussion as the classical pseudospin-Z expectation values are within $\lesssim 10^{-3}$ of being perfectly aligned/anti-aligned after the disappearance of the `island'.}
We have also used re-defined pseudospin-lattice couplings for brevity in the main text. 
These ``new'' couplings are related to the original phenomenological coupling constants:
$\mathcal{G}_z \equiv \frac{\left(g_{10} + 2g_9 \right)}{27 c_B}$, 
$\mathcal{G}_x \equiv  \frac{\left(g_{4} + 2g_3 \right)}{27 c_B}$,
$\mathcal{W}_x \equiv \frac{4\left(8\sqrt{2} g_1- 4g_2 \right)}{27 c_{44}} $,
$\mathcal{W}_z \equiv \frac{4 \left(8\sqrt{2} g_7- 4g_8 \right)}{27 c_{44}}$,
$\mathcal{V}_x \equiv \frac{4\left(g_{4} - g_3 \right)} {27 c_{44}} $,
$\mathcal{V}_z \equiv \frac{4\left(g_{10} - g_9 \right)}{27 c_{44}} $.

\section{Magnetostriction expressions along magnetic field $\bm{\hat{n}}$ and length change $\bm{\ell}$ directions}
\label{app_length_change_expressions}

We present here the complete magnetostriction expressions for the various experimentally relevant magnetic field and length change cubic directions.

\subsection{$\bf{H} \ ||$ [111]}
We repeat the expression for the $\left( \frac{\Delta L}{L} \right)^{[111]}_{(1,1,1)}$ magnetostriction in terms of the original pseudospin-elastic couplings for completeness.

\begin{widetext}
\begin{align}
\left( \frac{\Delta L}{L} \right)^{[111]}_{(1,1,1)} &= \frac{h}{27 c_B}  \left[  \left(g_{10} + 2g_9 \right)\left( 3 \tz0 - \tz1 - \tz2 - \tz3 \right) + \left(g_{4} + 2g_3 \right)\left( 3 \tx0 - \tx1 - \tx2 - \tx3 \right)  \right] \nonumber \\
& + \frac{4}{27 c_{44}} h  \Bigg[  \left(8\sqrt{2} g_1- 4g_2 \right) \left(\tx1 + \tx2 + \tx3 \right) + \left(g_4 - g_3 \right) \left(9 \tx0 + \tx1 + \tx2 + \tx3 \right) \Bigg.  \nonumber \\
&  \Bigg. ~~~~~~~~~~~~  + \left(8\sqrt{2} g_7- 4g_8 \right) \left(\tz1 + \tz2 + \tz3 \right) + \left(g_{10} - g_9 \right) \left(9 \tz0 + \tz1 + \tz2 + \tz3 \right) \Bigg]
\end{align}

\begin{align}
\left( \frac{\Delta L}{L} \right)^{[111]}_{(1,1,0)} &= \frac{h}{27 c_B} \left[  \left(g_{10} + 2g_9 \right)\left( 3 \tz0 - \tz1 - \tz2 - \tz3 \right) + \left(g_{4} + 2g_3 \right)\left( 3 \tx0 - \tx1 - \tx2 - \tx3 \right)  \right] \nonumber \\
& + \frac{1}{18 \sqrt{3} (c_{11} - c_{22}) } h \Bigg[  \left(\sqrt{6} g_1+ \sqrt{3}g_2 \right) \left(\tx1 + \tx2 -2 \tx3 \right) + \left(\sqrt{6} g_7+ \sqrt{3}g_8 \right) \left(\tz1 + \tz2 -2 \tz3 \right) \Bigg] \nonumber \\
& - \frac{2}{9 c_{44}}h \Bigg[ \left(g_3 - g_4 \right) \left(3\tx0 + \tx1 + \tx2  - \tx3 \right) + \left(g_2 - 2 \sqrt{2} g_1\right)\left(\tx1 + \tx2 + 2\tx3 \right) \Bigg. \nonumber \\
& ~~~~ ~~~~ ~ +  \left(g_9 - g_{10} \right) \left(3\tz0 + \tz1 + \tz2  - \tz3 \right) + \left(g_8 - 2 \sqrt{2} g_7\right)\left(\tz1 + \tz2 + 2\tz3 \right) \Bigg] \nonumber\\
& +  \Bigg[ \frac{\left(\sqrt{2} g_5- g_6 \right)}{6 \sqrt{3} (c_{11} - c_{22}) } + \frac{2 \left( 2 \sqrt{6} g_5 + \sqrt{3} g_6 \right)}{9 c_{44}}   \Bigg] h \left(\ty1 - \ty2 \right) 
\end{align}

\begin{align}
\left( \frac{\Delta L}{L} \right)^{[111]}_{(0,0,1)} &= \frac{h}{27 c_B}  \left[  \left(g_{10} + 2g_9 \right)\left( 3 \tz0 - \tz1 - \tz2 - \tz3 \right) + \left(g_{4} + 2g_3 \right)\left( 3 \tx0 - \tx1 - \tx2 - \tx3 \right)  \right] \nonumber \\
& - \frac{1}{9 \sqrt{3} (c_{11} - c_{22}) } h \Bigg[  \left(\sqrt{6} g_1+ \sqrt{3}g_2 \right) \left(\tx1 + \tx2 -2 \tx3 \right) + \left(\sqrt{6} g_7+ \sqrt{3}g_8 \right) \left(\tz1 + \tz2 -2 \tz3 \right) \Bigg. \nonumber \\
& \Bigg. ~~~~~~~~~~~~~~~~~~~~~~~~~ + \left(3 \sqrt{2} g_5 - 3g_6 \right) \left( \ty1 - \ty2 \right) \Bigg]
\end{align}
\end{widetext}

\subsection{$\bf{H} \ ||$ [110]}

\begin{widetext}
\begin{align}
\left( \frac{\Delta L}{L} \right)^{[110]}_{(1,1,1)} &= \frac{\sqrt{2}}{9 \sqrt{3} c_B} h \left[  \left(g_{10} + 2g_9 \right)\left( \tz0 -  \tz3 \right) + \left(g_{4} + 2g_3 \right)\left( \tx0 -  \tx3 \right)  \right] \nonumber \\
& + \frac{2}{27 c_{44}  } h \Bigg[  -2\sqrt{6} \left(g_3 - g_4 \right)\left(3 \tx0 + \tx3 \right) +  \sqrt{3} \left(4 g_1- \sqrt{2}g_2 \right) \left(3\tx1 + 3\tx2 + 2 \tx3 \right) \Bigg. \nonumber \\
& \Bigg. ~~ ~~ ~~ ~~ ~~ ~~ ~~ -2\sqrt{6} \left(g_9 - g_{10} \right)\left(3 \tz0 + \tz3 \right) +  \sqrt{3} \left(4 g_7- \sqrt{2}g_8 \right) \left(3\tz1 + 3\tz2 + 2 \tz3 \right) \nonumber \\
& \Bigg. ~~ ~~ ~~ ~~ ~~ ~~ ~~ + \left(12 g_5 + 3\sqrt{2} g_6 \right) \left( \ty1 - \ty2 \right) \Bigg]
\end{align}

\begin{align}
\left( \frac{\Delta L}{L} \right)^{[110]}_{(1,1,0)} &= \frac{\sqrt{2}}{9 \sqrt{3} c_B} h \left[  \left(g_{10} + 2g_9 \right)\left( \tz0 -  \tz3 \right) + \left(g_{4} + 2g_3 \right)\left( \tx0 -  \tx3 \right)  \right] \nonumber \\
& + \frac{1}{12 \sqrt{3} (c_{11} - c_{22})  } h \Bigg[ \left(2g_1 + \sqrt{2} g_2 \right) \left( \tx0 - \tx3 \right) + \left(2g_7 + \sqrt{2} g_8 \right) \left( \tz0 - \tz3 \right) \Bigg. \nonumber \\
& ~~~~~~~ ~~~~~~~~~~~~~ ~~~~~ +  \Bigg. \left(2 \sqrt{3} g_5 - \sqrt{6} g_6 \right) \left( \ty1 - \ty2 \right) \Bigg]  \\
& +\frac{1}{9 c_{44}} h \Bigg[ \sqrt{3} \left(-4g_1 + \sqrt{2} g_2 - 2\sqrt{2} g_3 + 2 \sqrt{2} g_4 \right)\left( \tx0 - \tx3 \right) \Bigg. \nonumber \\
& ~~~~~~ ~~+ \Bigg. \sqrt{3} \left(-4g_7 + \sqrt{2} g_8 - 2\sqrt{2} g_9 + 2 \sqrt{2} g_{10} \right)\left( \tz0 - \tz3 \right) + \left(12 g_5 + 3\sqrt{2} g_6 \right) \left( \ty1 - \ty2 \right) \Bigg] \nonumber
\end{align}

\begin{align}
\left( \frac{\Delta L}{L} \right)^{[110]}_{(0,0,1)} &= \frac{\sqrt{2}}{9 \sqrt{3} c_B} h \left[  \left(g_{10} + 2g_9 \right)\left( \tz0 -  \tz3 \right) + \left(g_{4} + 2g_3 \right)\left( \tx0 -  \tx3 \right)  \right] \nonumber \\
& + \frac{1}{6 \sqrt{3} (c_{11} - c_{22})  } h \Bigg[ \left(-2g_1 - \sqrt{2} g_2 \right) \left( \tx0 - \tx3 \right) + \left(-2g_7 - \sqrt{2} g_8 \right) \left( \tz0 - \tz3 \right) \Bigg. \nonumber \\
& ~~~~~~ ~~ ~~~~ ~~~~ ~~~~ ~~ +  \Bigg. \left(-2 \sqrt{3} g_5 + \sqrt{6} g_6 \right) \left( \ty1 - \ty2 \right) \Bigg] 
\end{align}
\end{widetext}

\subsection{$\bf{H} \ ||$ 001}

\begin{widetext}

\begin{align}
\left( \frac{\Delta L}{L} \right)^{[001]}_{(1,1,1)} &= \frac{1}{9 \sqrt{3} c_B}h  \left[  \left( 2g_3 + g_{4} \right)\left( \tx0 - \tx1 - \tx2  + \tx3 \right) + \left( 2g_9 + g_{10} \right)\left( \tz0 - \tz1 - \tz2  + \tz3 \right)  \right] \nonumber \\
& - \frac{4}{27 c_{44}} h  \Bigg[  \left(\sqrt{3} g_3- \sqrt{3} g_4 \right) \left(3 \tx0 + \tx1 + \tx2 - \tx3 \right) - \left(2 \sqrt{6} g_1 - \sqrt{3} g_2 \right) \left( \tx1 + \tx2 + 2 \tx3 \right) \Bigg.  \nonumber \\
&  \Bigg. ~~~~~~~~~~~~  + \left(\sqrt{3} g_9- \sqrt{3} g_{10} \right) \left(3 \tz0 + \tz1 + \tz2 - \tz3 \right) - \left(2 \sqrt{6} g_7 - \sqrt{3} g_8 \right) \left( \tz1 + \tz2 + 2 \tz3 \right)  \Bigg. \nonumber \\
& \Bigg.  ~~~ ~~~~~~~~~+ \left( 6 \sqrt{2} g_5 + 3g_6 \right) \left(\ty1 - \ty2 \right) \Bigg]
\end{align}

\begin{align}
\left( \frac{\Delta L}{L} \right)^{[001]}_{(1,1,0)} &= \frac{1}{9 \sqrt{3} c_B}h  \left[  \left( 2g_3 + g_{4} \right)\left( \tx0 - \tx1 - \tx2  + \tx3 \right) + \left( 2g_9 + g_{10} \right)\left( \tz0 - \tz1 - \tz2  + \tz3 \right)  \right] \nonumber \\
& + \frac{1}{6 \sqrt{3} (c_{11} - c_{12})} h \Bigg[ \left(-\sqrt{2} g_1 - g_2 \right) \left( \tx0 - \tx1 - \tx2 + \tx3 \right)  \Bigg. \nonumber \\ 
& \Bigg. ~~~~~~~~ ~~~~~~~~~~~~~~~~+ \left(-\sqrt{2} g_7 - g_8 \right) \left( \tz0 - \tz1 - \tz2 + \tz3 \right) \Bigg] \nonumber \\
& - \frac{2}{3 \sqrt{3} c_{44}} h  \Bigg[  \left(-2 \sqrt{2} g_1 + g_2 + g_3 - g_4 \right) \left(  \tx0 + \tx1 + \tx2 + \tx3 \right) \Bigg. \nonumber \\
& \Bigg. ~~~~~~~~ ~~ ~~~ + \left(-2 \sqrt{2} g_7 + g_8 + g_9 - g_{10} \right) \left(  \tz0 + \tz1 + \tz2 + \tz3 \right) \Bigg]
\end{align}

\begin{align}
\left( \frac{\Delta L}{L} \right)^{[001]}_{(0,0,1)} &= \frac{1}{9 \sqrt{3} c_B}h  \left[  \left( 2g_3 + g_{4} \right)\left( \tx0 - \tx1 - \tx2  + \tx3 \right) + \left( 2g_9 + g_{10} \right)\left( \tz0 - \tz1 - \tz2  + \tz3 \right)  \right] \nonumber \\
& + \frac{1}{3 \sqrt{3} (c_{11} - c_{12})} h \Bigg[ \left(\sqrt{2} g_1 + g_2 \right) \left( \tx0 - \tx1 - \tx2 + \tx3 \right)  + \left(\sqrt{2} g_7 + g_8 \right) \left( \tz0 - \tz1 - \tz2 + \tz3 \right) \Bigg] 
\end{align}

\end{widetext}

\section{32-site ED O-QSI transitions}

\begin{figure}[h]
  \centering
  \includegraphics[width=\linewidth]{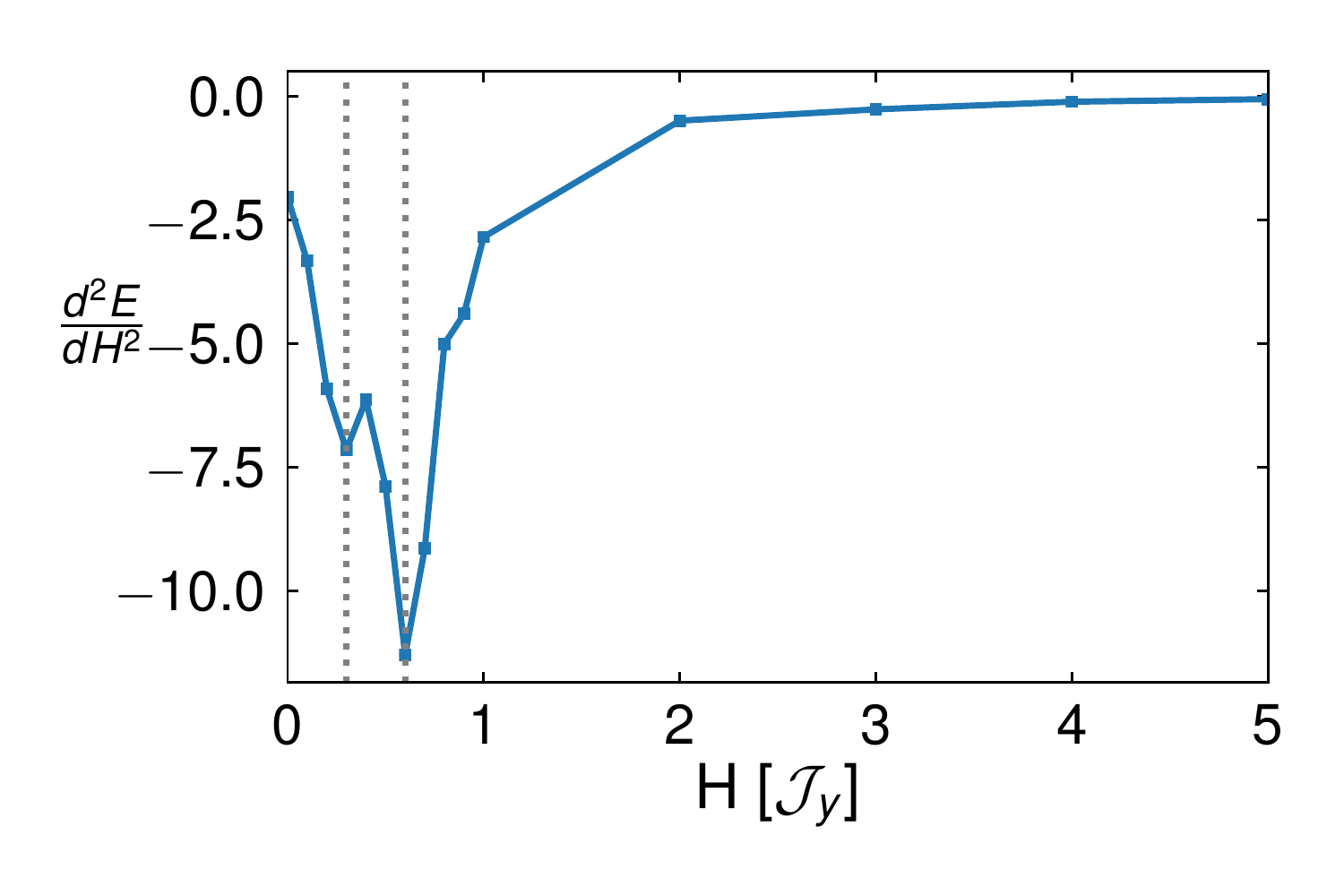}
  \caption{Second derivative of ED ground state energy (E) with respect to magnetic field strength for increasing magnetic field strength for O-QSI.
  There exist two discontinuous peaks at $h_{c1} \approx 0.3 \mathcal{J}_y$ and $h_{c2} \approx 0.6 \mathcal{J}_y$, indicating the second-order nature of the phase transition.
  $h_{c1} \approx 0.3 \mathcal{J}_{y}$ and $h_{c2} \approx 0.6 \mathcal{J}_{y}$ are indicated by grey-dotted lines.}
   \label{fig_ed_energy}
\end{figure}

We present in Fig. \ref{fig_ed_energy} the second derivative of the ED ground state energy as a function of magnetic field.
As seen (by the dashed vertical lines in Fig. \ref{fig_ed_energy}), there are two sharp dips, which indicate continuous (second) order phase transitions at $h_{c1} \approx 0.3 \mathcal{J}_y$ and $h_{c2} \approx 0.6 \mathcal{J}_y$.
\textcolor{black}{It is, however, unclear from the ED study as to whether these dips are due to finite size effects, or reflect true continuous phase transition points in the thermodynamic limit.}

\section{Magnetostriction of D-QSI and O-QSI along other cubic directions}
\label{app_qsi_110_001}

\begin{figure*}[t]
  \centering
  \includegraphics[width=0.88\linewidth]{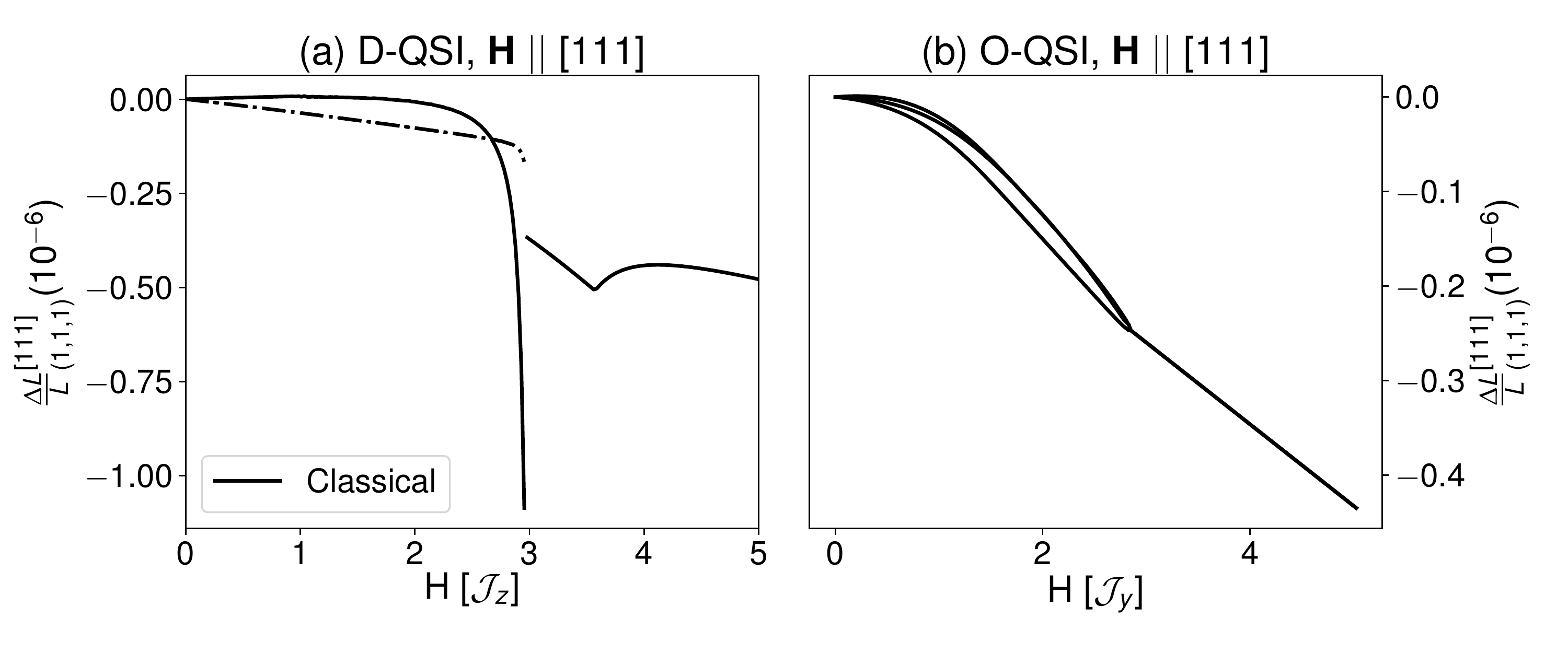}
  \caption{Length change, $\frac{\Delta L}{L}$, along $\bm{\ell} = (1,1,0)$ direction for magnetic field applied along $\bm{\hat{n}} = [111]$ direction for (a) dipolar quantum spin ice (D-QSI) and (b) octupolar quantum spin ice (O-QSI).
  Solid lines indicate classical magnetostrictions. 
  (a): D-QSI reflecting the Kagome ice degeneracy, and (b) O-QSI reflecting the `dampened' degeneracy of the O-QSI, as described in the main text.
  We denote the degenerate D-QSI solutions using dashed lines for ease of viewing.}
   \label{fig_qsi_110}
\end{figure*}

\begin{figure*}[t]
  \centering
  \includegraphics[width=0.88\linewidth]{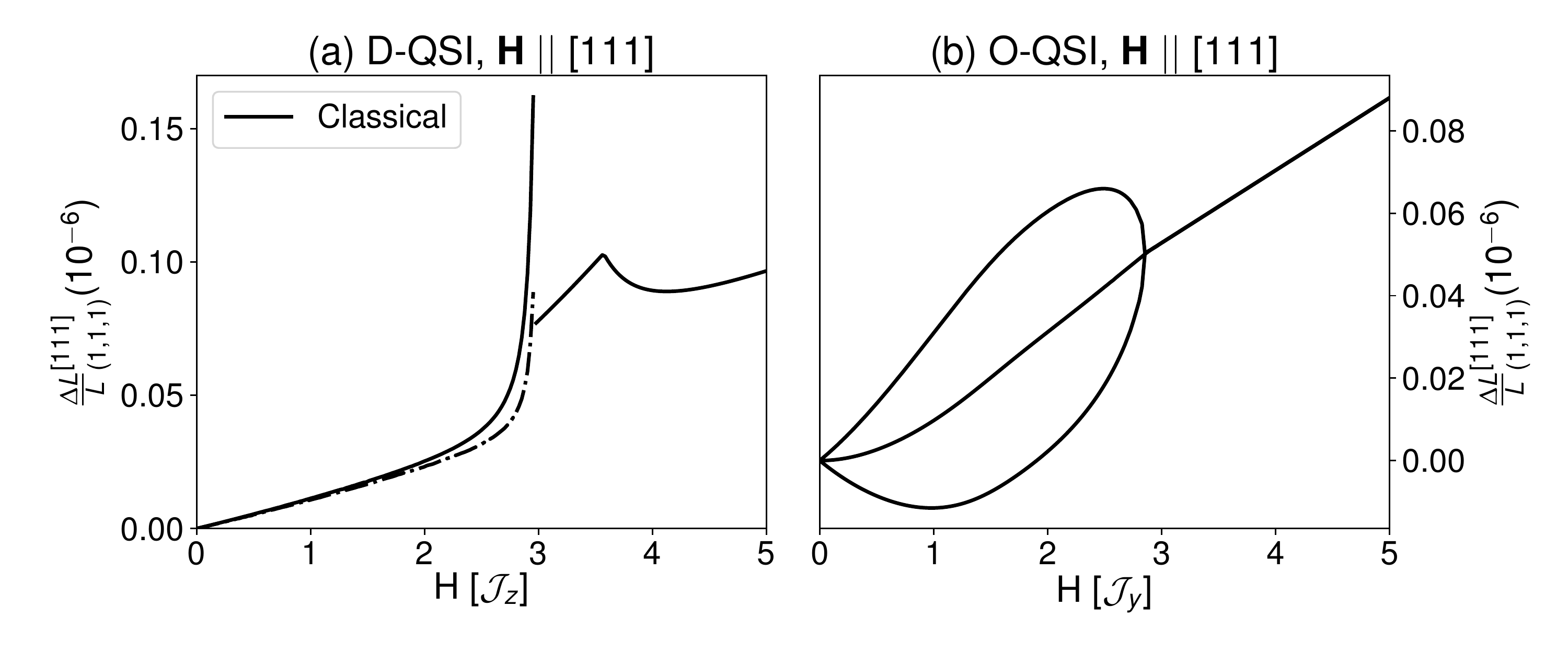}
  \caption{Length change, $\frac{\Delta L}{L}$, along $\bm{\ell} = (0,0,1)$ direction for magnetic field applied along $\bm{\hat{n}} = [111]$ direction for (a) dipolar quantum spin ice (D-QSI) and (b) octupolar quantum spin ice (O-QSI).
  Solid lines (squares) indicate classical (16-site exact diagonalization) magnetostrictions. 
  (a): D-QSI reflecting the Kagome ice degeneracy, and (b) O-QSI reflecting the `dampened' degeneracy of the O-QSI, as described in the main text.
  We denote the degenerate D-QSI solutions using dashed lines for ease of viewing.}  
   \label{fig_qsi_001}
\end{figure*}

We present in Figs. \ref{fig_qsi_110} and \ref{fig_qsi_001} the classical magnetostriction behaviours for the two quantum spin ices along the $\bm{\hat{n}} = [111]$ direction for length changes along the ${\bm{\ell}} = (1,1,0), (0,0,1)$ directions.
The presented directions provide the clearest differences between the D-QSI and O-QSI phases.
As seen, the degeneracy of the Kagome ice regime for D-QSI and the `dampened' degeneracy of the O-QSI are reflected in the classical solutions.
Since the [111] magnetic field does not couple to the $\tau^y$ octupolar moment, we are unable to extract out the ED pseudospin-Y expectation values needed for the ${\bf{\ell}} = (1,1,0), (0,0,1)$ directions, and so we only present the classical solutions.

%


\section{Magnetostriction of all-in, all-out multipolar ordered phases}
\begin{figure*}[h!]
  \centering
  \includegraphics[width=\linewidth]{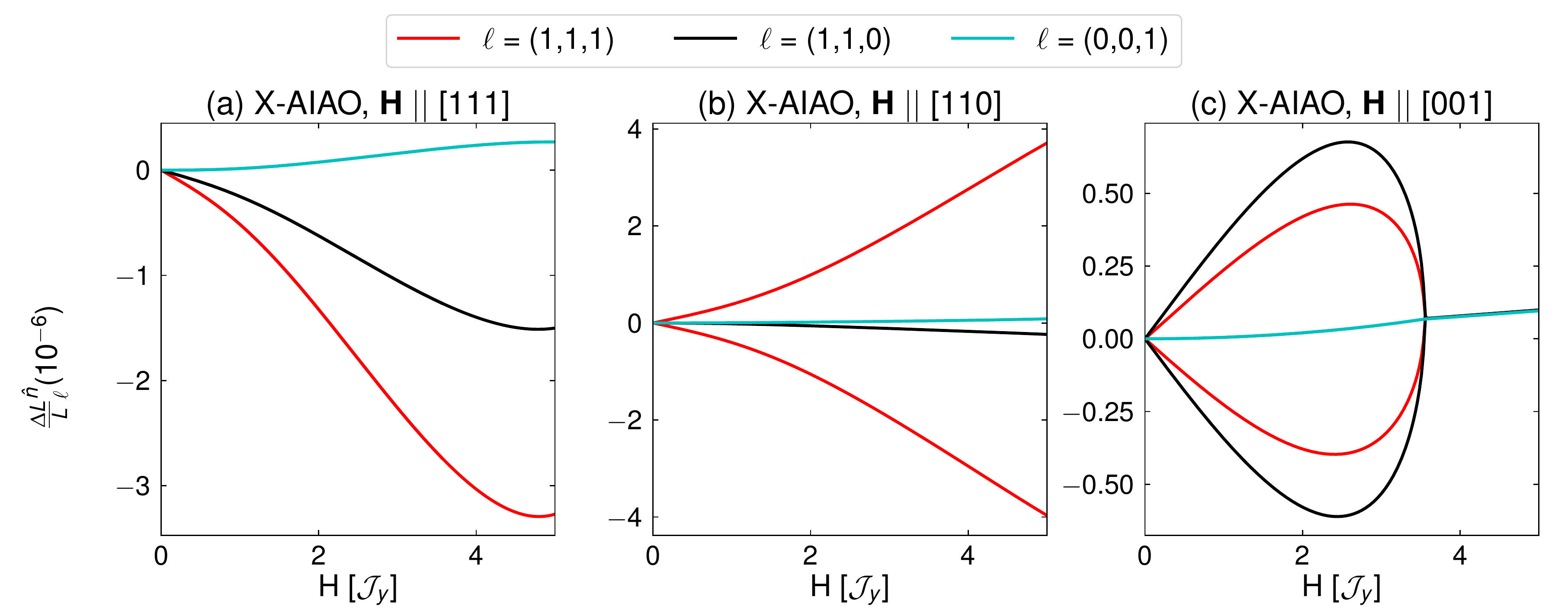}
  \caption{Classical magnetostriction behaviours, $\frac{\Delta L}{L}$, for magnetic fields applied along $\bm{\hat{n}}$ = [111], [110], [001] directions for X-AIAO phase $(\mathcal{J}_{x}, \mathcal{J}_{y}, \mathcal{J}_{z}) = (-0.75, 1, 0.2)$. 
Depicted are the three common experimentally accessible cubic length change directions $\bm{\ell} = (1,1,1), (1,1,0), (0,0,1)$ directions, in red, black and cyan, respectively.
The AIAO nature of the X-pseudospin results in the classically degenerate length change branches.
In a realistic system (with multiple domains), one expects an average over the degenerate branches.}
   \label{fig_x_aiao_all}
\end{figure*}

\begin{figure*}[h]
  \centering
  \includegraphics[width=1\linewidth]{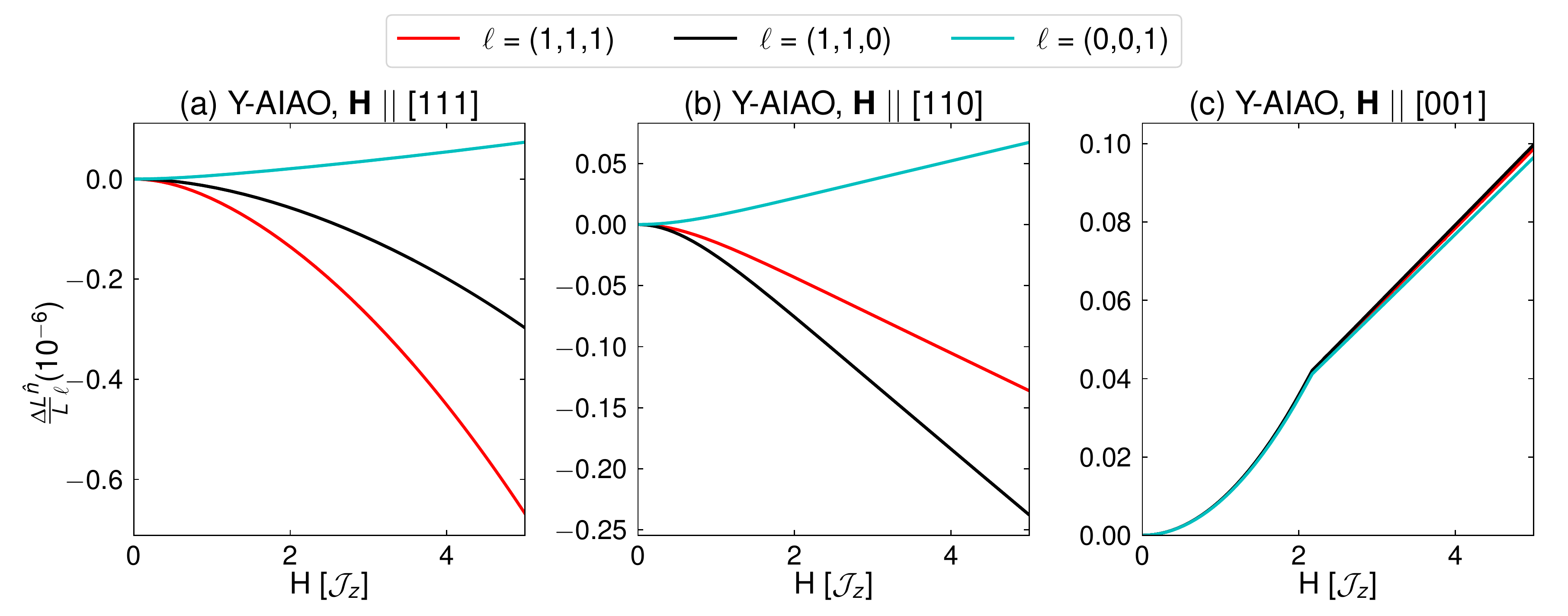}
  \caption{Classical magnetostriction behaviours, $\frac{\Delta L}{L}$, for magnetic fields applied along $\bm{\hat{n}}$ = [111], [110], [001] directions for Y-AIAO phase $(\mathcal{J}_{x}, \mathcal{J}_{y}, \mathcal{J}_{z}) = (0.2, -0.75, 1)$. 
Depicted are the three common experimentally accessible cubic length change directions $\bm{\ell} = (1,1,1), (1,1,0), (0,0,1)$ directions, in red, black and cyan, respectively.
The AIAO nature of the Y-pseudospin is not reflected as a multiple branches, due to the lack of particular combinations of $\tau^y$ in the length change to highlight the classical degeneracy.
In a realistic system (with multiple domains), one expects an average over the degenerate branches.}
   \label{fig_y_aiao_all}
\end{figure*}

\begin{figure*}[t]
  \centering
  \includegraphics[width=1\linewidth]{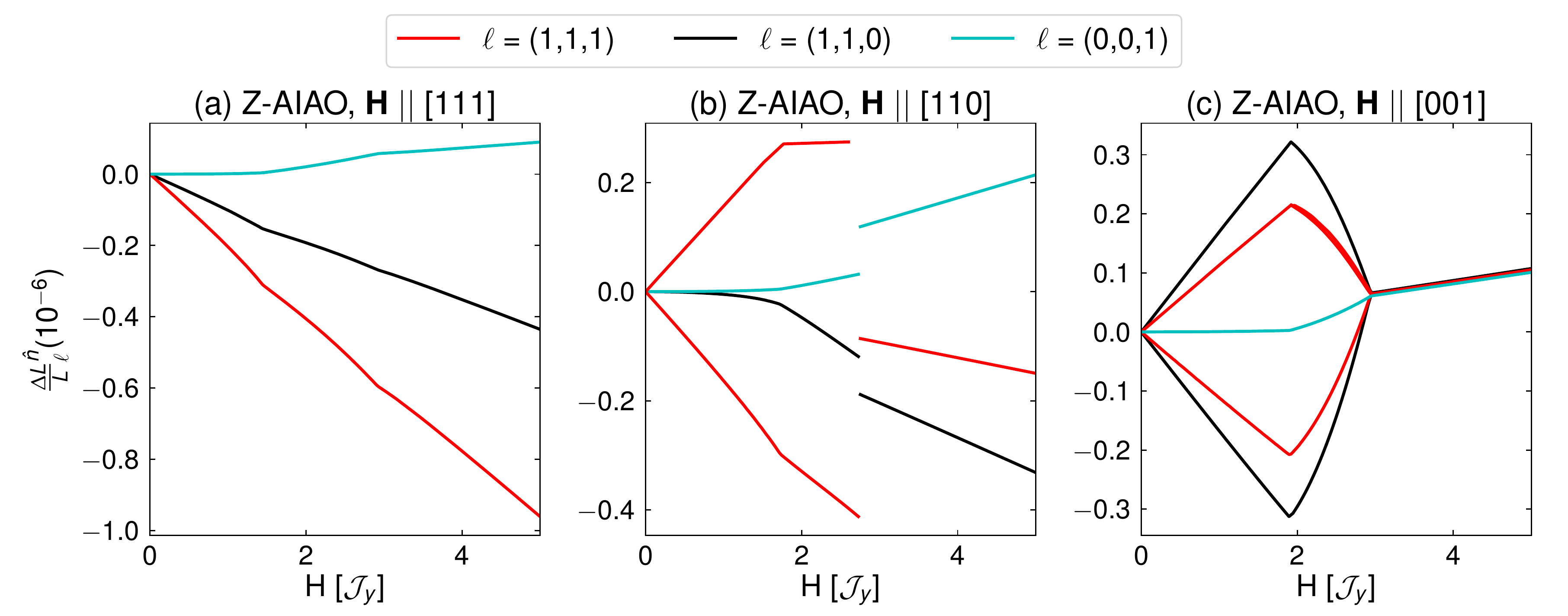}
  \caption{Classical magnetostriction behaviours, $\frac{\Delta L}{L}$, for magnetic fields applied along $\bm{\hat{n}}$ = [111], [110], [001] directions for Z-AIAO phase $(\mathcal{J}_{x}, \mathcal{J}_{y}, \mathcal{J}_{z}) = (-0.1, 1, -0.7)$. 
Depicted are the three experimentally accessible cubic length change directions $\bm{\ell} = (1,1,1), (1,1,0), (0,0,1)$ directions, in red, black and cyan, respectively.
There exist classically degenerate branches at low and intermediate fields, reflecting the degeneracy of the all-in, all-out nature of the Z-AIAO phase.
In a realistic system (with multiple domains), one expects an average over the degenerate branches.}
   \label{fig_o_z_aiao_all}
\end{figure*}

The DO spin ice phases are flanked by multipolar ordered all-in, all-out (AIAO) phases where the expectation values of the pseudospin operators on each sublattice is the same: X-AIAO ($\langle \tau^x _{\alpha} \rangle= m_x$), Y-AIAO ($\langle \tau^y _{\alpha} \rangle= m_y$), and Z-AIAO ($\langle \tau^z _{\alpha} \rangle= m_z$) $\forall \alpha$, where $\{m_x,m_y,m_z\} \in \mathbb{R}^3$.
We present the magnetostriction of these AIAIO phases under $\bf{H} \ ||$ [111], [110] and [001] in Figs. \ref{fig_x_aiao_all}, \ref{fig_y_aiao_all}, and \ref{fig_o_z_aiao_all}.
We note that there exist degenerate branches for the AIAO magnetostriction behaviours, which reflects the degeneracy of the AIAO phase (i.e. $m_x>0$ or $m_x<0$ etc.).
Clearly, this is not observed in all the length change directions, as it requires particular combinations of the pseudospin configuration to appear in the length change expressions such as $\sim h (3 \tz1 + 3 \tz2 + 2\tz3)$, $\sim h (3 \tz0 + \tz3) $, $\sim h (3 \tz1 + 3 \tz2 + 2\tz3)$, $\sim h (3 \tz0 + \tz3)$, and $\sim h (\ty1 - \ty2) $.
Nevertheless there are many possible behaviours (continuous, non-analytic `kinks'), which highlights the anisotropic nature of magnetostriction, and offers explicit selection rules to identify the ordered phases.

\clearpage
\bibliography{do_bibtex}

\end{document}